\documentclass[12pt]{article}
\usepackage{amsmath,latexsym}
\usepackage{graphicx}
\begin{document}

 \title{\Huge Wormhole with varying cosmological constant}
 \author{F.Rahaman$^*$, M.Kalam$^{\ddag}$, M.Sarker$^*$,
 A.Ghosh$^*$and B.Raychaudhuri$^\dag$}

\date{}
 \maketitle
 \begin{abstract}
It has been suggested that the cosmological constant is a
variable dynamical quantity. A class of solution has been
presented for the spherically symmetric space time describing
wormholes by assuming the erstwhile cosmological constant
$\Lambda$ to be a space variable scalar, viz., $\Lambda $ =
$\Lambda (r) $ . It is shown that the Averaged Null Energy
Condition (ANEC) violating exotic matter can be made arbitrarily
small.
\end{abstract}

  \footnotetext{ Pacs Nos :  04.20 Gz,04.50 + h, 04.20 Jb   
 $*$Dept.of Mathematics, Jadavpur University, Kolkata-700 032, India
                                  E-Mail:farook\_rahaman@yahoo.com\\
$\ddag$Dept. of Phys. , Netaji Nagar College for Women, Regent Estate, Kolkata-700092, India.\\
  $\dag$Dept. of Phys. , Surya Sen Mahavidyalaya, Siliguri, West Bengal, India}
In 1917, Einstein introduced cosmological constant $\Lambda$ which
is related to the energy of the space to maintain the stability of
his cosmological model. Recent observations of high redshift Type
$I_a$ supernovae ~\cite{kg1} suggest that $\Lambda$ could be
scalar variable dependent both on space and time coordinate rather
a constant, which was believed earlier. Time dependence of
$\Lambda$ plays a significant role on cosmology whereas space
dependence of $\Lambda$ does effect on astrophysical problems.
Narlikar et al~\cite{kg2} have suggested that space dependence of
$\Lambda$ should be included to study the nature of local massive
objects like galaxies. Recently, some authors have shown
significant effects of the space dependence $\Lambda$ on energy
density of the classical electron~\cite{kg3}. Tiwari et
al~\cite{kg4} and Dymnikova~\cite{kg5} have discussed the
contribution of space dependence $\Lambda$ to the effective
gravitational mass of the astrophysical system.

Wormholes are classical or quantum solutions for the gravitational
field equations describing a bridge between two asymptotic
manifolds. Classically, they can be interpreted as instantons
describing a tunneling between two distant regions. In a pioneer
work, Morris and Thorne [6] have shown that a wormhole geometry
can only appear as a solution to the Einstein equation if the
stress energy tensor violates the null energy condition. The
matter that characterized stress energy tensor is known as exotic
matter [7]. Several authors have discussed wormholes in scalar
tensor theory of gravity in which scalar field may play the role
of exotic matter[8]. Recently, authors are interested to know how
much exotic matter is needed to get a traversable wormhole [9].
Peebles and Ratra [10] proposed that like all energy, cosmological
constant $\Lambda$ has some contribution to the source term in
Einstein's gravitational field equations. It is believed through
indirect evidences that 70 percent of the contents of the Universe
is to be in the form of vacuum energy
 and cosmological constant turns to be a measure of the energy density of the vacuum.
 In an interesting paper Lemos et al have studied wormhole
 geometry in presence of  $\Lambda$  where  $\Lambda$  is a
 constant[11].
  In this work, we are interested to discuss and provide a prescription for obtaining
  wormhole solution by
assuming cosmological constant $\Lambda$ to be a space variable
scalar. To our knowledge, wormhole solution under the assumption
that the cosmological constant is spatially variable has not been
proposed earlier.

The Einstein field equation for the anisotropic fluid
distribution are given by

\begin{equation} \label{Eq1}
 R_{ab} -  \frac {1}{2}  g_{ab} +  \Lambda  g_{ab}   =   - 8 \pi T_{ab} \text{( $G = c = 1$ )}.
  \end{equation}
where matter momentum tensor is given by $T_a^b = ( - \rho, p,
p_t, p_t )$ and the related conservation law here is
\begin{equation}\label{Eq2}
8 \pi [T_a^b]_{;b} =  - \Lambda _{;b}
\end{equation}
as the cosmological constant is assumed to be spatially varying
i.e. $\Lambda $ = $\Lambda (r) $.

[ The usual energy momentum tensor is modified by the addition of
a term $ T_{ab} ^ {(vac)} = - \Lambda(r) g_{ab}$ . Hence, the new
energy momentum tensor is $ T_{ab} ^ {(total)} = T_{ab} ^ {(
matter ) } + \Lambda(r) g_{ab}$. Here energy conservation equation
$ { T_a^b} ^{(total)}_{;b} = 0 $ implies equation (2)  ]

 Let us  now consider the spherically symmetric line element

\begin{equation}\label{Eq3}
ds^2 = - e^\gamma dt^2 + e^\mu dr^2 + r^2 d\Omega_2^2,
\end{equation}
where
\begin{equation}\label{Eq4}
e^{-\mu} = \left[1 - \frac{b(r)}{r}\right].
\end{equation}

Here $\displaystyle{\frac{\gamma(r)}{2}}$ is the redshift
function and $b(r)$ is the shape function determining the shape
of the wormhole.

The field equations~\eqref{Eq1} corresponding to the above line
element~\eqref{Eq3} are given by

\begin{equation}\label{Eq5}
 e^{-\mu}\left[ - \frac{1}{r^2}+\frac{\mu^\prime}{r}\right]+\frac{1}{r^2}
 =  8\pi \rho + \Lambda,
 \end{equation}

 \begin{equation}\label{Eq6}
  e^{-\mu}\left[\frac{1}{r^2} +\frac{\gamma^\prime}{r}\right]-\frac{1}{r^2}
  =  8\pi p - \Lambda,
 \end{equation}

\begin{equation}\label{Eq7}
  \frac{1}{2}e^{-\mu}\left[\gamma^{\prime\prime} +
  \frac{1}{2}(\gamma^\prime)^2-\frac{1}{2}\gamma^\prime \mu^\prime +  \frac{\gamma^\prime -
  \mu^\prime}{r}\right] = 8\pi p_t - \Lambda,
 \end{equation}

where $p$, $ p_t$ are  radial and tangential pressures
respectively and $\rho$ is the matter energy density.

 [`$\prime$' refers to differentiation  with respect to radial coordinate.]

The conservation equation~\eqref{Eq2} becomes,

 \begin{equation}\label{Eq8}
  \frac{d}{dr}\left( p - \frac{\Lambda}{8\pi}\right) =
  - ( p + \rho)\frac{\gamma\prime}{2}+\frac{2}{r}
  ( p_t  - p).
  \end{equation}

In this work, we are not interested in discussing the
traversability constrains mentioned by Morris and
Thorne~\cite{kg6}. We assume a zero tidal force as seen by the
stationary observer, $\frac{\gamma(r)}{2} = 0$, to make the
problem simpler. We suppose also that the pressures are
anisotropic and

\begin{equation}\label{Eq9}
 p_t = n p.
\end{equation}

( $n$ is an arbitrary constant )

Now,  from the field equation~\eqref{Eq6}, one finds,

\begin{equation}\label{Eq10}
e^{-\mu} = \left[1 - \frac{b(r)}{r}\right]= 1 + 8\pi p r^2
-\Lambda r^2,
\end{equation}

where

\begin{equation}\label{Eq11}
b(r) = \Lambda r^3  - 8\pi p r^3.
 \end{equation}

Equation~\eqref{Eq8} yields

\begin{equation}\label{Eq12}
  \frac{d}{dr}\left( p - \frac{\Lambda}{8\pi}\right) = \frac{2( n - 1 )}{r}p.
 \end{equation}

Since the vacuum energy ( which is equivalent to $\Lambda$) can be
thought as a contributor of the anisotropic fluid distribution, we
impose the condition, $ \frac{\Lambda}{8\pi} \propto p$ for
simplicity and this implies
\begin{equation}\label{Eq13}
\frac{\Lambda}{8\pi} = a p.
\end{equation}
[ $a$ is the proportional constant ]

By solving Eq.~\eqref{Eq12}, one obtains,

\begin{equation}\label{Eq14}
p = A r ^{-B},
\end{equation}

 where

\begin{equation}\label{Eq15}
 B =  \frac{ 2(n - 1)}{( a - 1 )},
 \end{equation}

 and $A$ is the  integration constant.

 The expression for $b(r)$ is

\begin{equation}\label{Eq16}
 b(r) = 8\pi A( a - 1 )r ^{ (- B + 3) },
 \end{equation}

where $a > 1$ .

Since the space time is asymptotically flat i.e.
$\frac{b(r)}{r}\rightarrow 0 $ as $ \mid r \mid \rightarrow
\infty $, the Eq.~\eqref{Eq16} is consistent only when $B > 2 $
i.e. $n > a> 1$.

Eq.~\eqref{Eq5}, after some rearrangement, reduces to

\begin{equation}\label{Eq17}
\begin{aligned}
 \rho &= A( B - 3 - aB + 2a )r ^{- B}\\
    \end{aligned}
 \end{equation}

Using eqs. (14) and (17), one can find that
\begin{equation}
 p + \rho = - 2( n - a )A r ^{- B} < 0
 \end{equation}
 since, $ n > a > 1 $.

 Thus, null energy condition is violated.

 Now, we will check whether the wormhole geometry in principle,
 supported by arbitrary amount of Averaged Null Energy
Condition (ANEC) violating exotic matter. The ANEC violating
matter can be quantified by the integrals [9]
\begin{equation} I =
\oint ( p_i + \rho ) dV  \end{equation}
 In this model, we assume
that the ANEC violating matter is related only to p ( radial
pressure) , not to the transverse components [ as one can see
from eqs. (9), (14) and (17), that the sign of $ p_t + \rho $ is
not fixed but depends on the values of the parameters ].

Now, if one assumes, $ n = a$  + so called $ \epsilon $, then $ n
- a =  \epsilon $ , in other words, the integral (19) tends to
zero as $ \epsilon \rightarrow  0 $. Hence the ANEC violating
matter can be made arbitrarily small.

 The throat of the wormhole occurs at

\begin{equation}\label{Eq20}
r_0 =  [ 8\pi A ( a - 1) ]^{\frac{1}{ (B - 2)} }.
\end{equation}

The axially symmetric embedded surface $ z = z(r)$ shaping the
Wormhole's spatial geometry is a solution of
\begin{equation}\label{Eq21}
 \frac{dz}{dr}=\pm \frac{1}{\sqrt{\displaystyle{\frac{r}{b(r)}}-1}}.
 \end{equation}
  One can note from the definition of Wormhole that at   $ r= r_0 $
  (the wormhole throat) Eq.~\eqref{Eq21} is divergent i.e.  embedded surface is
   vertical there.

The embedded surface (solution of Eq.~\eqref{Eq21}) in this case
is [ we assume $ B = 4 $ ],

\begin{equation}\label{Eq22}
z =  \sqrt{8\pi A ( a - 1)} \cosh^{-1}   \frac{r}{\sqrt{8\pi A (
a - 1)}}.
\end{equation}

\begin{figure}[htbp]
    \centering
        \includegraphics[scale=.8]{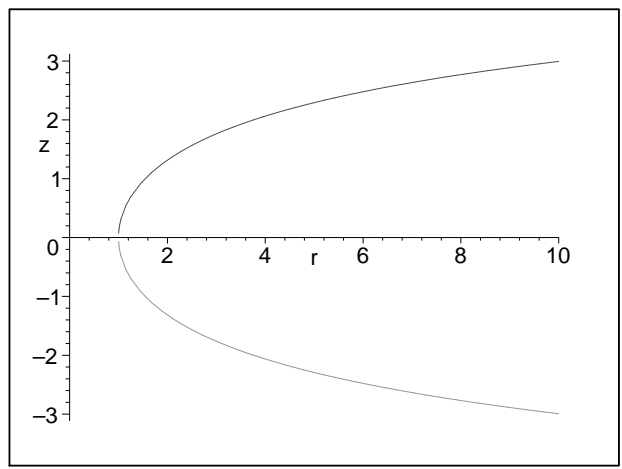}
        \caption{The embedding diagram of the wormhole}
    \label{fig:cosh}
\end{figure}

\begin{figure}[htbp]
    \centering
        \includegraphics[scale=.8]{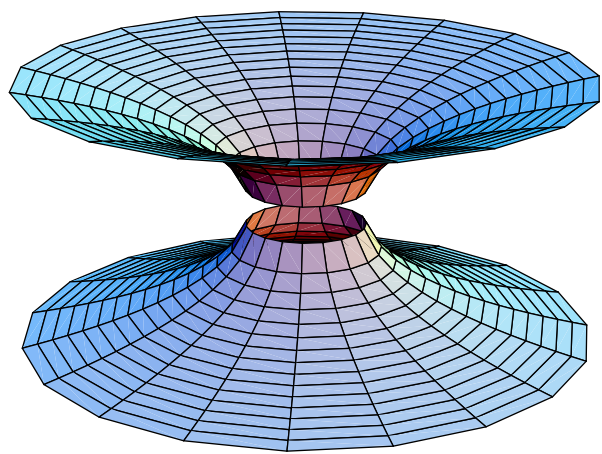}
    \caption{The full visualization of the surface generated by the rotation of the embedded
    curve about the vertical z axis .}
    \label{fig:wormhole}
\end{figure}

One can see embedding diagram of this wormhole in
Fig.~\ref{fig:cosh}. The surface of revolution of this curve
about the vertical z axis makes the diagram complete
(Fig.~\ref{fig:wormhole}).

\pagebreak

According to Morris and Thorne~\cite{kg6} , the '$r$' co-ordinate
is ill-behaved near the throat, but proper radial distance
\begin{equation}\label{Eq23}
 l(r) = \pm \int_{r_0^+}^r \frac{dr}{\sqrt{1-\frac{b(r)}{r}}}
 \end{equation}
 must be well behaved everywhere i.e. we must require that $ l(r) $is finite throughout the space-time.

 In this model ( for  $B = 4$  ),

\begin{equation}\label{Eq24}
l(r) =  \pm\sqrt {r^2 - 8 \pi A ( a - 1) }.
 \end{equation}

Due to the simple expression for $l(r)$, one can rewrite the
metric tensor in terms of this proper radial distance,

\begin{equation}\label{Eq25}
  ds^2 = -  dt^2 + dl^2 + r^2(l) d\Omega_2^2,
  \end{equation}

where

\begin{equation}\label{Eq26}
r^2(l) = l^2 + 8 \pi A ( a - 1).
\end{equation}

This is a well behaved coordinate system. The radial distance is
positive above the throat (our Universe) and negative below the
throat (other Universe). At very large distance from the throat,
the embedding surface becomes flat $\displaystyle{\frac{dz}{dr}}(
l \longrightarrow  \pm  \infty ) = 0 $ corresponding to  the two
asymptotically flat regions ($l \longrightarrow+\infty $ and $l
\longrightarrow- \infty$), which the wormhole connects.

In conclusion, our aim in this work has been to provide a
prescription for obtaining wormhole in presence of variable
cosmological constant. The most striking features of our model is
that if we choose the parameters,  'n' is very close to 'a', then
ANEC violating matter can be made arbitrarily small. Our wormhole
can be visualized by the surface of revolution of the curve $r =
\sqrt{8\pi A ( a - 1)}
\cosh\left[\displaystyle{\frac{z}{\sqrt{8\pi A ( a -
1)}}}\right]$. Though this research work is mostly theoretical in
nature, the outcome of the  result may be of interest to the
researchers working in this field. The traversable wormhole opens
up several possible interesting physical applications and we hope
to report on this elsewhere.

\pagebreak

 { \bf Acknowledgments }

          F.R. is thankful to Jadavpur University and DST , Government of India for providing
          financial support under Potential Excellence and Young
          Scientist scheme . MK has been partially supported by
          UGC,
          Government of India under Minor Research Project scheme.
          We are also grateful to the  referees for their valuable
comments and constructive suggestions.
          \\

\end{document}